\documentclass{elsart}

\begin{document}
\begin{frontmatter}
\title{ Universal properties of conformal quantum  many-body systems}
\author{Stjepan Meljanac}
\address{Rudjer Bo\v skovi\'c Institute, Bijeni\v cka  c.54, 
          HR-10002 Zagreb, Croatia}
\ead{meljanac@irb.hr}

\author{Andjelo Samsarov}
\address{ Rudjer Bo\v skovi\'c Institute, Bijeni\v cka  c.54, 
          HR-10002 Zagreb, Croatia}
\ead{asamsarov@irb.hr}

\begin{abstract}
  Universal properties of many-body systems in conformal quantum mechanics in
arbitrary dimensions are presented.
Specially, a general structure of discrete energy spectra and eigenstates is found.
Finally, a simple construction of a 
universal time operator conjugated to a conformal Hermitian or a $ PT-$ invariant
Hamiltonian is proposed.

\end{abstract}
\end{frontmatter}


\newpage



A number of physical systems exhibit a particular form of asymptotic conformal
invariance. There are examples from 
black holes to molecular and condensed matter physics that can be
 discussed within a unified treatment. The symmetries relevant here
include time translation invariance, scale invariance, and invariance under special
 conformal transformations that are all part of a larger conformal symmetry with $
SO(2,1) $
 group-theory structure [1].

 A connection of conformal symmetry with black hole physics has recently been
explored in various forms. In particular,
the near-horizon symmetry structure of black holes 
as well as the impact of this symmetry on thermodynamics [2] of black holes has been
considered together with its
extension to superconformal quantum mechanics [3,4].
 In the near-horizon 
area of black holes, the geometry is that of the three-dimensional anti-de Sitter
gravity and, by the
Henneaux-Brown conjecture [5], can be described with the help of  conformal field
theory [6].
By imposing suitable boundary conditions at the horizon, it can be shown that the
actual algebra of surface deformations
contains a Virasoro algebra in the $ ( r-t ) $ plane. Another route of current
applications involves the link of
black holes to the Calogero model [7,8], which originates from the fact that the
dynamics of particles 
near the horizon of a black hole is associated with a Hamiltonian containing an
inverse square potential
 [9,10] and conformal symmetry [1]. It is therefore of interest
to find quantum states of systems in conformal quantum mechanics, that, in the
context of black holes appear
to be the horizon states. 
 Some attempts in this direction have been made in various papers [10,11].

In molecular physics, conformal invariance is important in the context of the
occurrence of anomaly that is revealed by
the failure of symmetry generators to close the algebra. It is shown that the
concept of anomaly applies to the inverse 
square potential, including the electric dipole-charge interaction, with strong
implications in
 molecular physics [12].
Conformal quantum mechanics is also important in condensed matter physics.
For example, a correlated  
two-dimensional $ N- $ electron gas with an inverse-square interaction in a magnetic
field has been studied
 within a microscopic analytical theory [13].
In this letter we present simple and universal results for many-body systems in
conformal quantum mechanics
in arbitrary dimensions.



Let us consider  $ N $ generally different particles in arbitrary dimensions $ D $,
described by a  
Hermitian or a $ PT- $ invariant Hamiltonian [14] of the form
\begin{equation}
  H = - \frac{1}{2}\sum_{i=1}^{N}\frac{1}{m_{i}} {{\nabla}_{i}}^{2}
         + V( \vec{r}_{i},...,  \vec{r}_{N} ) +
\frac{{\omega}^{2}}{2}\sum_{i=1}^{N}m_{i} {\vec{r}_{i}}^{2} +
   \frac{c}{2} ( \sum_{i=1}^{N} {\vec{r}_{i}} {{\nabla}_{i}} + \frac{ND}{2}).
\end{equation}
The potential $ V $ is a real homogeneous function or a $ PT- $ invariant operator
of order
 $ -2 $, i.e., it satisfies the relation
\begin{equation}
  [ \sum_{i=1}^{N} {\vec{r}_{i}} {{\nabla}_{i}}, V ]  = -2V . 
\end{equation}
Additionally, we assume that $ V $ is invariant under translations, i.e., 
$ [ \sum_{i=1}^{N} {{\nabla}_{i}}, V ]  = 0 \; $ and, generally, it can be
unisotropic [12].
Then the Hamiltonian (1) can be written as $ \;\; H = -T_{-} + {\omega}^{2}T_{+} +
cT_{0} \;\; $, where the
generators  $\{T_{\pm},T_0\}$, defined as
\begin{equation}\begin{array}{l}
 T_{+} = \frac{1}{2}\sum_{i=1}^{N}m_{i} {\vec{r}_{i}}^{2}, \\ 
 T_{-} = \frac{1}{2}\sum_{i=1}^{N}\frac{1}{m_{i}} {{\nabla}_{i}}^{2}
         - V( \vec{r}_{i},...,  \vec{r}_{N} ), \\
 T_{0} = \frac{1}{4}  \sum_{i=1}^{N} ({\vec{r}_{i}} {{\nabla}_{i}} + {{\nabla}_{i}} \vec{r}_{i})  =
 \frac{1}{2} \sum_{i=1}^{N} {\vec{r}_{i}} {{\nabla}_{i}} + \frac{ND}{4},
\end{array}\end{equation}
 satisfy the SU(1,1) conformal algebra
\begin{equation}
[T_{-},T_{+}] = 2T_{0}, \;\;\;\; [T_{0},T_{\pm}] = \pm T_{\pm}.
\end{equation}
The generators $ T_{\pm} $ ( acting on the Hilbert space of physical states)
are Hermitian operators, whereas $ T_{0} $ is an anti-Hermitian operator. If $ V $
is not Hermitian but
 $ PT- $ invariant , so is the  $ T_{-} $ generator.

The above system can be viewed as a deformation of $ N $ harmonic oscillators in $ D
$ dimensions,
 with a common frequency $ \omega $.
 The Hamiltonian (1) can be mapped to $ \;\; 2 \omega'T_{0} \;\; $
by  transformation:
\begin{equation}
 H = 2 \omega' S T_{0} S^{-1},
\end{equation}
where
\begin{equation}
    S = e^{- b T_{+}} e^{- a T_{-}}
\end{equation}
  and
\begin{equation}\begin{array}{l}
 a = \frac{ 1}{2 \omega} \frac{1 }{ \sqrt{1 + \frac{c^{2}}{4 {\omega}^{2} } }}, \\
 b = \omega ( \sqrt{ 1 +  \frac{ c^{2}}{4 {\omega}^{2}}} - \frac{c}{2 \omega} ), \\
  \omega' = \omega \sqrt{ 1 +  \frac{ c^{2}}{4 {\omega}^{2}}}.
\end{array}\end{equation}
If $ c $ is real, the new frequency $ \omega' $ is greater than the original
frequency $ \omega $, and 
the Hamiltonian (1) is $ PT- $ invariant. If
$ c = 0 $, then $ \omega' = \omega $. If $ c $ is pure imaginary, then $ \omega' $
is less than $ \omega $ 
and the $ \; cT_{0} \; $ part, as well as the whole Hamiltonian, is Hermitian.
However, for $ c = \pm 2 \imath \omega $, it holds $ \omega' = 0 $, the 
$ S $ transformation becomes singular, and the system is critical. This point represents a boundary 
between the region of discrete energies which occurs for $ \; {\omega'}^{2} > 0 \; $ and the continuum energy
spectrum describing the scattering states that appears for $ \; {\omega'}^{2} < 0. $ The case $ \; {\omega'}^{2} < 0 \; $
is related to the notion of inverted oscillator which is relevant to string theory.
Note that the norms of the wave functions blow up at the point $ \; \omega' = 0  $.

The ground state $ \psi_{0} $ is well defined if it is a square integrable function.
This will be the case
 if the ground-state energy $ \; \omega' \epsilon_{0}, \;\;  \epsilon_{0} > 0, \;  $
 is higher than $ \; \omega' \frac{D}{2}, $ the condition which is connected with  the existence of
the critical point [15,16].
 Then we can write $ \;\; \psi_{0} = S \Delta_{0}, \;\; $ where 
$ \Delta_{0} $ is a homogeneous function of the lowest degree and of the lowest energy
 $ \omega'  \epsilon_{0}, $  satisfying
\begin{equation}
T_{-} \Delta_{0} = 0,  \;\;\;\;  T_{0} \Delta_{0} =  \frac{\epsilon_{0}}{2} \Delta_{0}.
\end{equation}
There are other homogeneous, generally irrational, solutions $ \Delta_{k} $ with
homogeneity of higher degrees
 and with energy $ \omega' \epsilon_{k} = \omega' ( \epsilon_{0} + k ), \; k > 0  $
 (generally, there are degenerate states with the same $k $ ).
They satisfy
\begin{equation}
  T_{-} \Delta_{k} = 0,  \;\; \;\; T_{0} \Delta_{k} =  \frac{ k + \epsilon_{0}}{2}
\Delta_{k},  \;\; k > 0.
\end{equation}
Since there is a class of $ PT- $ invariant Hamiltonians that have  real energy
spectra [14],
we are free to  restrict ourselves to the cases where the energies  $ \omega'
\epsilon_{k} $ of the system (1) are real. 
For example, if $ V = 0 $, then $ \Delta_{k} $ are harmonic polynomials in $ ND $
dimensions
with energies $ \; \omega' ( \frac{ND}{2} + k ), \; $ where $ k = 0,1,2,....$
The construction of  solutions of Eqs. (9) with $ V \neq 0, $ is generally a
difficult task. 

The universal set of excited states is $  \;\; \psi_{n,k} = S {T_{+}}^{n}
\Delta_{k},   \;\; n = 0,1,2,...; \;
 k \geq 0, \;\; $ with energies  $ \;\; 2\omega' ( n +  \frac{ \epsilon_{k}}{2} ) 
\;\; $. For a given
$ k \geq 0, $ the spectrum is equidistant with an elementary step $  2\omega'  $.
Hence, all solutions are grouped into equidistant towers based on $ S \Delta_{k},
\;\; k \geq 0 $ .
 Specially, for identical particles in one dimension,  the states 
$  \;\; \psi_{n,k} = S {T_{+}}^{n} \Delta_{k},   \;\; n, k = 0,1,2,...,  \;\; $
represent the complete set of 
 $ S_{N} $ symmetric solutions, with an equidistant energy spectrum with the step $
\; \omega' $.

Note that there is a universal radial equation for a radial part $ \;
\phi_{n,k}(T_{+}) \; $ of the wave function
 $ \; \psi_{n,k} = \phi_{n,k}(T_{+}) \Delta_{k} \; $.
Using the radial representation of the generators  $ \; T_{\pm}, T_{0}, \; $ with $
T_{+} $ as a radial variable,
Eq.(3), and with
\begin{equation}
T_{-} = T_{+} \frac{{d}^{2}}{d{T_{+}}^{2}} +  \epsilon_{k} \frac{d}{d T_{+}}  ,
  \;\;\;\;  T_{0} =  T_{+} \frac{d}{d T_{+}}  +  \frac{\epsilon_{k}}{2},
\end{equation}
then by applying them to the $ S \Delta_{k} $ tower, one obtains the universal
radial equation 
\begin{equation}
 [  T_{+}\frac{{d}^{2}}{d{T_{+}}^{2}} + ( \epsilon_{k} - cT_{+} )
    \frac{d }{d T_{+}} + ( E_{n,k} - { \omega}^{2} T_{+} - c \frac{ \epsilon_{k}}{2} )]
  \phi_{n,k}(T_{+}) = 0.
\end{equation}
It follows directly upon substituting the factorization $ \; \psi_{n,k} =
\phi_{n,k}(T_{+}) \Delta_{k} \; $
into the eigenvalue equation  
 $ \;\; H \psi_{n,k}  = E_{n,k} \psi_{n,k}, \;\; $
with solutions $ \; \psi_{n,k} = S {T_{+}}^{n} \Delta_{k}, \; n = 0,1,2,... $  and
with the corresponding energies
 $ \; E_{n,k} = \omega' ( 2n + \epsilon_{k})  $.
The solution of Eq. (11) is of the form $ \;  \phi_{n,k}(T_{+}) = F_{n,k}(T_{+}) e^{
- \omega' T_{+}}, \; $ where,
 $  \; F_{n,k}(T_{+}) \; $ is an associated Laguerre polynomial $ \; 
L_{ n + \epsilon_{k} - 1}^{\epsilon_{k} - 1} ( 2 \omega T_{+}) \; $ for $ \; c = 0 $.  
The set of operators $ \{ T_{\pm}, T_{0} \} $ can be separated into the
center-of-mass and relative parts:
$ \; T_{\pm} = (T_{CM})_{\pm} + (T_{rel})_{\pm} \; $  and $ \; T_{0} = (T_{CM})_{0}
+ (T_{rel})_{0} \; $ with  
the same separation for the energy parameter $ \epsilon_{k} $. The universal
radial equation then splits into two equations of the form (11), namely, for
 $ \; (T_{CM})_{+} \; $ and for $ \; (T_{rel})_{+} \; $.
Generally, the center-of-mass $(CM) $ motion is described by the $ D- $ dimensional
oscillator, see Refs. [15,16].
Then the excited states can be written as 
$ \psi_{ n_{\alpha}, n, k } = S \prod_{\alpha = 1}^{D} {(
R_{\alpha})}^{n_{\alpha}}(T_{rel})_{+}^{n} \Delta_{k}, \;\; $
 where $ \;\; \vec{R} = \frac{ \sum_{i = 1 }^{N} m_{i} {\vec{r}}_{i}} {\sum_{i = 1
}^{N} m_{i}} \;\; $
 is the center-of-mass vector.
 
For example, for an arbitrary conformal quantum many-body system in two dimensions
with identical particles [8,13,16],
 the corresponding states are
\begin{equation}
 S ( \sum_{i = 1}^{N} z_{i} )^{l}  ( \sum_{i = 1}^{N} \bar{ z}_{i} )^{\bar{l}}
(T_{rel})_{+}^{n} \Delta_{k} \;,
\end{equation}
where  $ \; z_{i}, {\bar{z}}_{i} \; $ are complex coordinates. These states are the
eigenstates of the angular momentum
 operator, with the eigenvalue $ \;\; L_{CM} = l - \bar{l} \;\; $ due to the $ CM $
motion.

 The Fock space corresponding to
  $  \;\; \psi_{n,k} = S {T_{+}}^{n} \Delta_{k},   \;\; n = 0,1,2,...; \;
 k \geq 0, \;\; $ is spanned by 
 $  \;\; \prod_{\alpha = 1}^{D} {({A_{1, \alpha}}^{+})}^{n_{\alpha}}
 {({B_{2}}^{+})}^{n} \psi_{0,k}, \;\; $ where 
$  \;\;  \psi_{0,k} = S \Delta_{k} \;\; $ are vacua of corresponding towers. We have
introduced the operators
$ \;\; {A_{1, \alpha}}^{\pm}   \; $ and $ \; {B_{2}}^{\pm} \;\; $ defined by
\begin{equation}\begin{array}{l}
  {A_{1, \alpha}}^{+} = S R_{ \alpha} S^{-1}, \qquad  {A_{1, \alpha}}^{-} = S 
{\nabla}_{ \alpha} S^{-1}, \\
  {B_{2}}^{\pm} = S { ( T_{rel})}_{\pm} S^{-1}.
\end{array}\end{equation}

Furthermore, we construct universal phase-angle variables and a time operator
conjugated to the Hamiltonian (1),
generalizing the result of Ref. [17]. From the
$ SU(1,1) $ algebra it follows 
\begin{equation}\begin{array}{l}
  T_{+}T_{-} = \phi ( T_{0}) = ( T_{0} - \frac{ \epsilon }{2}) (  T_{0} + \frac{
\epsilon }{2} - 1 ), \\
  T_{-}T_{+} = \phi ( T_{0} + 1) = ( T_{0} - \frac{ \epsilon }{2} + 1) (  T_{0} +
\frac{ \epsilon }{2}),
\end{array}\end{equation}
where $ \omega' \epsilon $ is the generic energy of the lowest state in the
corresponding tower $ \; $ ( cf. Eq. (9)).
Also, for an arbitrary function $ f( T_{0} ) $ which has the power series expansion,
it holds
$  \;\; T_{-}f( T_{0} ) = f( T_{0} + 1 )T_{-} \;\; $ and $  \;\; T_{+}f( T_{0} ) =
f( T_{0} - 1 )T_{+} \;\; $.
The Casimir operator is $ \; -T_{+}T_{-} + T_{0}(T_{0} - 1 ) \; $ with the eigenvalue
$ \; \frac{ \epsilon_{k}}{2} ( \frac{ \epsilon_{k}}{2} - 1 ) \; $ when acting on the
particular tower based on
$ \Delta_{k} $. The quantity $ \; \frac{ \epsilon_{k}}{2} > 0 \; $ is a spin of a
discrete irreducible representation 
of the universal covering group [18].

Now we can introduce the operator
\begin{equation}
  Q = T_{+}\frac{\imath }{T_{0} + \frac{ \epsilon }{2} } 
\end{equation}
which is conjugated to  $ \;  -T_{-} \;  $ for each tower of states built on $
\Delta_{k} $ (for $ \; \epsilon > 0, \; $ the energy spectrum is discrete).
 Then the following relations hold: 
\begin{equation}
 [ Q, -T_{-}] = \imath, \;\;\;\;  [ Q, T_{+}] = \imath Q^{2}, \;\;\;\; [ Q, T_{0}] = -Q.
\end{equation}
It is easy to verify that the adjoint form of the relations (16) looks like
$$
 [ Q^{\dagger}, -T_{-}] = \imath, \;\;\;\;  [ Q^{\dagger}, T_{+}] = \imath {Q^{\dagger}}^{2}, \;\;\;\;
  [ Q^{\dagger}, T_{0}] = -Q^{\dagger}.
$$
Since $ Q $ is not a Hermitian operator, we define a Hermitian one as
 $ {\mathcal{T}}_{0}  =  \frac{1}{2}( Q + {Q}^{\dagger} ) $.
Now the operator conjugate to the Hamiltonian $ \;\; H = -T_{-} + {\omega}^{2}T_{+}
+ cT_{0} \;\; $
  is obtained by solving the equation
\begin{equation}
 [ {\mathcal{F}}(Q), H ] = \imath.
\end{equation}
Anticipating the adjoint form of relations (16), note that this relation
 directly implies $ \; [ {\mathcal{F}}(Q^{\dagger}), H ] = \imath $.
By using Eqs. (16), the relation (17) is reduced to 
$$
 \frac{ d {\mathcal{F}}} { d Q} ( 1 + {\omega}^{2}{Q}^{2} + \imath cQ ) = 1
$$
and the result for the time operator $ \; {\mathcal{T}} \; $ is
\begin{equation}
 {\mathcal{T}} = \frac{1}{2}( {\mathcal{F}(Q)} + \mathcal{F}(Q^{\dagger})),
\end{equation}
where
\begin{equation}
 {\mathcal{F}(Q)} =  \frac{1 }{ \sqrt{1 + \frac{c^{2}}{4 {\omega}^{2} } }} \frac{ 1}{2 \omega}
    arctg \biggl( \frac{\omega Q + \frac{ \imath c }{2\omega } }{ \sqrt{1 +
\frac{c^{2}}{4 {\omega}^{2} } }} \biggr).   
\end{equation}
In the case that the Hamiltonian $ \; H \; $ is Hermitian, the time operator $ \; {\mathcal{T}}, \; $
Eq. (18), is also Hermitian.

The time operator can be written in terms of the logarithmic function by using the identity
$ \;  arctg x = \frac{ 1}{2 \imath} ln \frac{ 1 + \imath x}{ 1 - \imath x} \;  $ .  
 There are generally many time operators corresponding to towers
( generally, there are an infinite number of towers ) with different 
$ \;  \epsilon_{k} =  \epsilon_{0} + k, \;\; k \geq 0 $. Besides this, one can add
an arbitrary function of the 
Hamiltonian to  $ {\mathcal{T}} $, Eq.(18),  without changing anything.
One can also define the
 time operator in respect to the center-of-mass and relative Hamiltonians.
 Our result, Eq.(18), for the time operator
is a simple universal property of all systems with the underlying $ SU(1,1) $
symmetry .

In the case of one oscillator in $ D $ dimensions $ (V = 0) $, the operator $ Q, $
Eq.(15), reduces to the form (in the following we set $ \; m = 1 $ ).
\begin{equation}
 Q = {\vec{R}}^{2} {(\vec{P}\cdot \vec{R})}^{-1}, \;\;\;\; \vec{P} = -\imath \nabla .
\end{equation}
Particularly, the result (18) is a generalization of Ref. [17]. For one
 harmonic oscillator in one dimension with vanishing 
potential $ ( V = 0 ), $ we obtain 
\begin{equation}
 {\mathcal{T}} =  \frac{ 1}{2 \omega}
    arctg ( \omega x \frac{ 1}{p} ) + h.c.  ,
\end{equation}
coinciding with Ref. [17]. In the limiting case $ \; c = 0, \;\; \omega \rightarrow 0 \; $,
we obtain
$ \; {\mathcal{T}}  = {\mathcal{T}}_{0} = \frac{1}{2}( Q + {Q}^{\dagger} ) \; $ and
for $ \; V = 0, \; $ it reduces to the
time operator of the free particle 
\begin{equation}
 {\mathcal{T}} =  \frac{ 1}{2 } \biggl( x \frac{ 1}{p} + \frac{ 1}{p} x \biggr).
\end{equation}
This quantity coincides exactly with the time-of-arrival operator of Aharonov and Bohm [19].

Finally, one can introduce  the operators
\begin{equation}
 A_{2}^{\pm} = S {(2 \omega')}^{\pm 1} T_{\pm} S^{-1} =
    \frac{ { (\omega' \pm \frac{c}{2})}^{2}}{2 \omega'} T_{+} + \frac{1}{2 \omega'} T_{-}
          - ( \frac{c}{2 \omega'} \pm 1) T_{0},
\end{equation}
where $ \; \omega' \; $ is given in the relation (7) and $ \; c \; $ is the parameter appearing in front of the
$ \; PT- $ invariant term in the expression (1).
Owing to the relations $ \;\; [ H, {A_{2}}^{\pm} ] = \pm 2 \omega' {A_{2}}^{\pm},
\;\; $
 the time operator can be written as
\begin{equation}
 {\mathcal{T}} = - \frac{ \imath}{4\omega' } \biggl( ln {A_{2}}^{+} - ln {A_{2}}^{-}
\biggr).
\end{equation}
Note that the operator 
 $ \;\; \tilde{{\mathcal{T}}} = - \frac{ \imath}{4\omega' } \biggl( ln T_{+} - ln
T_{-} \biggr) \;\; $ 
is conjugated to  $ \;\; 2 \omega' T_{0},  \;\; $ i.e., $ \;\; [
\tilde{{\mathcal{T}}}, 2 \omega' T_{0}] = \imath,  \;\; $ 
where $ \;\; \tilde{{\mathcal{T}}} = S^{-1} {\mathcal{T}} S,  \;\; $ 
 and  $ S $ is the rotation operator defined by (6). 
 The coherent states can be constructed similarly as in Refs. [17,18].

In summary, we have presented some simple and
 universal properties of a general class of many-body systems in conformal quantum
mechanics in arbitrary 
dimensions. Particularly, the radial equation, common to all systems with underlying
conformal symmetry,
has been constructed, and a general structure of discrete energy spectra and
eigenstates has been found.
The universal form for the operator conjugated to the Hamiltonian has also been found.
  The methods presented here can be extended to $ N- $ body  systems in superconformal 
quantum mechanics.



This work was supported by the Ministry of Science and Technology of the Republic of
Croatia under 
contract No. 0098003.




\end{document}